\documentclass[twocolumn,showpacs,superscriptaddress,amssymb,10pt]{revtex4-1}

\usepackage{graphicx}
\usepackage{dcolumn}
\usepackage{bm}
\usepackage{float}
\usepackage{epsfig}
\usepackage{color}
\usepackage{longtable}

\definecolor{aogreen}{rgb}{0.0, 0.5, 0.0}

\usepackage{booktabs}
\newcommand{\bc}{\begin{color}}
\newcommand{\ec}{\end{color}}
\newcommand{\beqa}{\begin{eqnarray}}
\newcommand{\eeqa}{\end{eqnarray}}
\newcommand{\bit}{\begin{itemize}}
\newcommand{\eit}{\end{itemize}}
\newcommand{\ie}{\emph{i.e.}\ }
\newcommand{\beq}{\begin{equation}}
\newcommand{\eeq}{\end{equation}}
\newcommand{\twoline}[2][c]{%
  \begin{tabular}[#1]{@{}c@{}}#2\end{tabular}}

\begin{document}
\preprint{}
\title{Electronic isotope shift factors for the Cu $4s \; ^2S_{1/2} - 4p \; ^2P^o_{3/2}$ line}

\author{T. Carette}

\author{M. Godefroid}
\affiliation{Chimie quantique et photophysique, CP160/09, Universit\'e libre de Bruxelles, B 1050 Brussels, Belgium}

\date{\today \\[0.3cm]}

\begin{abstract}
   State-of-the-art relativistic multiconfiguration Dirac-Hartree-Fock calculations have been performed to evaluate the electronic field and mass isotope shift factors of the Cu~I resonance  line at $\lambda = 324.8$~nm. A linear correlation between the mass factors and the transition energy is found for elaborate correlation models, allowing extrapolation to the observed frequency limit. The relativistic corrections to the recoil operator reduces the transition mass factor by 5~\%.
\end{abstract}

%
%
\maketitle

\bigskip

\section{Introduction}

Nuclear mean-square charge radii  provide a highly sensitive test for nuclear shell effects~\cite{Kreetal:2014a}. Changes in the mean-square charge radius along a chain of isotopes can be deduced from the observed isotope shift in a given atomic transition $k$ of  frequency $\nu_k$.  The isotope shift depends on both the nuclear and electronic properties of the atom.  
Typically the observed isotope shift  $\delta \nu_k^{A,A'}$ between any pair of isotopes with masses $m_A$ and $m_{A'}$ is related to  the change in mean-square charge radius of the nuclear charge distribution between both isotopes, $\delta\langle r^2 \rangle^{A,A'}$, through the following expression:
\begin{equation}
\label{Line_IS}
 \delta \nu_k^{A,A'} \equiv \nu_k^{A'} - \nu_k^{A}  = M_k \; \left( \frac{m_{A'}-m_{A}}{m_{A}m_{A'}} \right)  + F_k \; \delta \langle r^2 \rangle^{A,A'} \; .
\end{equation}
$M_k$ and $F_k$ are respectively  the transition mass-shift and  field-shift electronic parameters.
In order to obtain these essential atomic data for the
Cu $4s \; ^2S_{1/2} - 4p \; ^2P^o_{3/2}$ resonance  at $\lambda = 324.8$~nm, state-of-the-art multi-configuration Dirac-Hartree-Fock (MCDHF) calculations have been performed for both levels, permitting a confident interpretation of the measured isotope shifts
~\cite{Bisetal:2016a}.

\section{Computational strategy and correlation models}

We use the multi-configuration Dirac-Hartree-Fock method~\cite{Gra:2007a} to compute highly accurate wave-functions for the
[Ar]$3d^{10}4s \; ^2S_{1/2}$ and
[Ar]$3d^{10}4p \; ^2P^o_{3/2}$ states of Cu~I.  
One way to determine  $M_k$ and $F_k$ appearing in Eq.~(\ref{Line_IS}) using an \emph{ab initio} method is to first compute the energies of the upper and lower atomic levels for several isotopes. This can be done by diagonalizing the full Hamiltonian matrix including the specific mass shifts (SMS)~\cite{JonFro:97a,Fri:2001a} and the extended nucleus charge distribution~\cite{Paretal:96a}.  For a given $(A,A',A'')$ triad, inverting the $(2 \times 2)$ system of equations from the line shifts (\ref{Line_IS}) yields the electronic parameters~\cite{Cheetal:2012a}.
Such calculations performed by Fritzsche are reported in~\cite{Bisetal:2016a}, restricting electron correlation to single and double excitations from the $3d$ and $4s$ shells into one or two layers of correlation orbitals and adding single excitations from the core to these correlation shells. These ``limited correlation model" (LCM) calculations are extended  in the present work, to provide reliable  estimations of the relevant electronic parameters. The relativistic corrections to the recoil operator~\cite{Sha:85a,Sha:88a}, that are neglected in~\cite{JonFro:97a,Fri:2001a}, are also investigated. While the relativistic recoil operator has been mostly used for a few electron systems \cite{Lietal:2012a,Zubetal:2014a,Nazetal:2014a}, studies remain scarce for neutral and singly ionized heavier atomic systems~\cite{KozKor:2007a,Nazetal:2015a}.

In this work, we report on large scale MCDHF calculations in which the expectation values of the relevant operators~\cite{Nazetal:2013a} are estimated using the $^{65}$Cu wave functions calculated with  GRASP2K~\cite{Jonetal:2013a}.
In this scheme, $M_k$ is the difference in the expectation values between the two levels of the recoil Hamiltonian containing both the one- and two-body terms constituting respectively  the NMS and SMS, as described in \cite{Nazetal:2013a}. In the case of Cu~I, $F_k$ can be reliably computed using the electron density at the nucleus~\cite{Ekm:2014a,Lietal:2012a}.
The impact of relativistic corrections to the recoil operator as originally derived by Shabaev~\cite{Sha:85a,Sha:88a} is estimated by comparing  the results obtained with \textsc{SMS92}~\cite{JonFro:97a}  (denoted $\overline{M}_k$), based on the Dirac kinetic form of the NMS~\cite{Lietal:2012b} and the mass polarisation term $\langle \frac{1}{M} \sum_{i<j} {\bf p}_i \cdot {\bf p}_j \rangle  $ for the SMS, with the values (denoted $M_k$), obtained with  \textsc{RIS3}~\cite{Nazetal:2013a} using the relativistically-corrected recoil operator~\cite{Gaietal:2011a}.
 
Although the considered electric dipole transition refers nominally to a rather simple single-electron excitation outside of a closed core $\mbox{[Ar]}~3d^{10}4s \rightarrow \mbox{[Ar]}~3d^{10}4p$, the interaction between the nearly degenerate $3d$ and $4s$ shells strongly affects this transition. Including single and double excitations of the main configuration up to numerical saturation of the variational space, recovers only 80\% of the total electron correlation contribution to the $4s_{1/2}$ -- $4p_{3/2}$ transition energy.

It becomes intractable to consider all triple and quadruple excitations of the main configuration in any reasonably large set of orbitals. An efficient way for capturing electron correlation is to pre-define a multi-reference (MR) space spanned by the most important configuration state functions (CSF) giving a good zeroth-order physical picture of the desired state. A multi-reference interacting (MR-I) expansion is then built~\cite{Caretal:2010a} as a CSF expansion
\beqa \label{eq:MR-I}
\Psi(\gamma\ J\pi)
= \sum_i c_i \Phi_i(\gamma_i\ J\pi) \; ,
\eeqa
containing all $\Phi_i(\gamma_i\ J\pi)$ that can be produced for a given orbital active set~(AS) 
\[ \{\phi_{n \kappa} \hspace*{0.2cm}; \hspace*{0.2cm} n<n_{max}\; , \;  l<l_{max} \}, \]
with the restriction that they  interact to first order with at least one of the reference states of the MR space \ie
\beqa\label{eq:MR-I}
&\Phi_i(\gamma_i\ J\pi) \in \text{ MR-I } \Leftrightarrow \exists \;  \Phi_k( \gamma_k\ J\pi) \in\text{ MR }  &\nonumber\\
& \text{ with } \left\langle\Phi_i(\gamma_i\ J\pi) \left| H \right| \Phi_k(\gamma_k\ J\pi) \right\rangle \neq 0\quad \forall~\{\phi_{n \kappa}\} \; .
\eeqa
Here, $H$ is the Dirac-Coulomb Hamiltonian~\cite{Gra:2007a}. The condition $\forall~\{\phi_{n \kappa}\} $ in (\ref{eq:MR-I}) excludes accidental zeros in the interaction matrix elements and guarantees their occurrence as being due to the spin-angular algebra.
 \emph{Ab initio} predictions can then be supported by a series of results obtained by extending systematically the orbital active set  on the one hand, and the multi-reference on the other hand.

Standard MCDHF calculations on medium-sized systems like Cu~I require to target the model on the desired quantity. Here, not all CSFs interacting with the MR in Eq.~(\ref{eq:MR-I}) are included, but rather they are limited to excitations of specific subshells.
Reminding that the two main configurations are of the type [Ar]$3d^{10}n\ell$, with $(n\ell)=4s/4p$, we will distinguish three types of double excitations:
- core-core (CC), involving $n=2,3$ electrons; - core-valence (CV), involving $n=3$ and $4s/4p$ electrons and - inner-core-valence (ICV), involving $1s^2$ and $4s/4p$ electrons.
All calculations are performed with a common orbital basis for the $4s$ and $4p$ states, \ie  ``Extended Optimized Level'' (EOL) type of calculations~\cite{Graetal:80a} on the two states. The AS is extended layer by layer, and only the last layer is optimized at each subsequent MCDHF calculation.
We begin with a two configuration Dirac-Fock (DF) calculation on the $4s$ and $4p$ states. Orbitals are optimized with only the main configurations in the MR. In all calculations $4d$ and $4f$ correlation orbitals are omitted. We optimize the $n=5$ layer on core-core correlation (CC). In a first series of calculations, we add core-valence (CC+CV) excitations, extending the AS until convergence (up to $n_{max}l_{max}= 10h$).
In a second series of calculations, we add core-valence and inner-core-valence excitations (CC+CV+ICV). For these, we reach a satisfactory convergence at $n_{max}l_{max}=12h$.

Where the $4s$ and $4p$ states differ significantly is in the high order correlation effects implying the $3d$ and valence electrons. Therefore, we merge the model omitting inner-core-valence with the \mbox{MR-I} set built by keeping the Ar-like core closed  and including one by one
$\{3d^94s6d, 3d^94p6f, 3d^96p6f\}$ and $\{3d^94s6p, 3d^94s4p, 3d^94s6f \}$ for the $4s$ state and  $4p$ state, respectively.
Note that the $n=6$ layer is the first core-valence correlation layer, the $n=5$ orbitals being optimised specifically on CC correlation. The resulting lists are used in multi-reference relativistic configuration interaction (RCI) calculations, including Breit interaction and vacuum-polarization.

The numerically converging results are summarized in Table~\ref{tab:res} for the different calculations.

\begin{figure}
\includegraphics[scale=0.25]{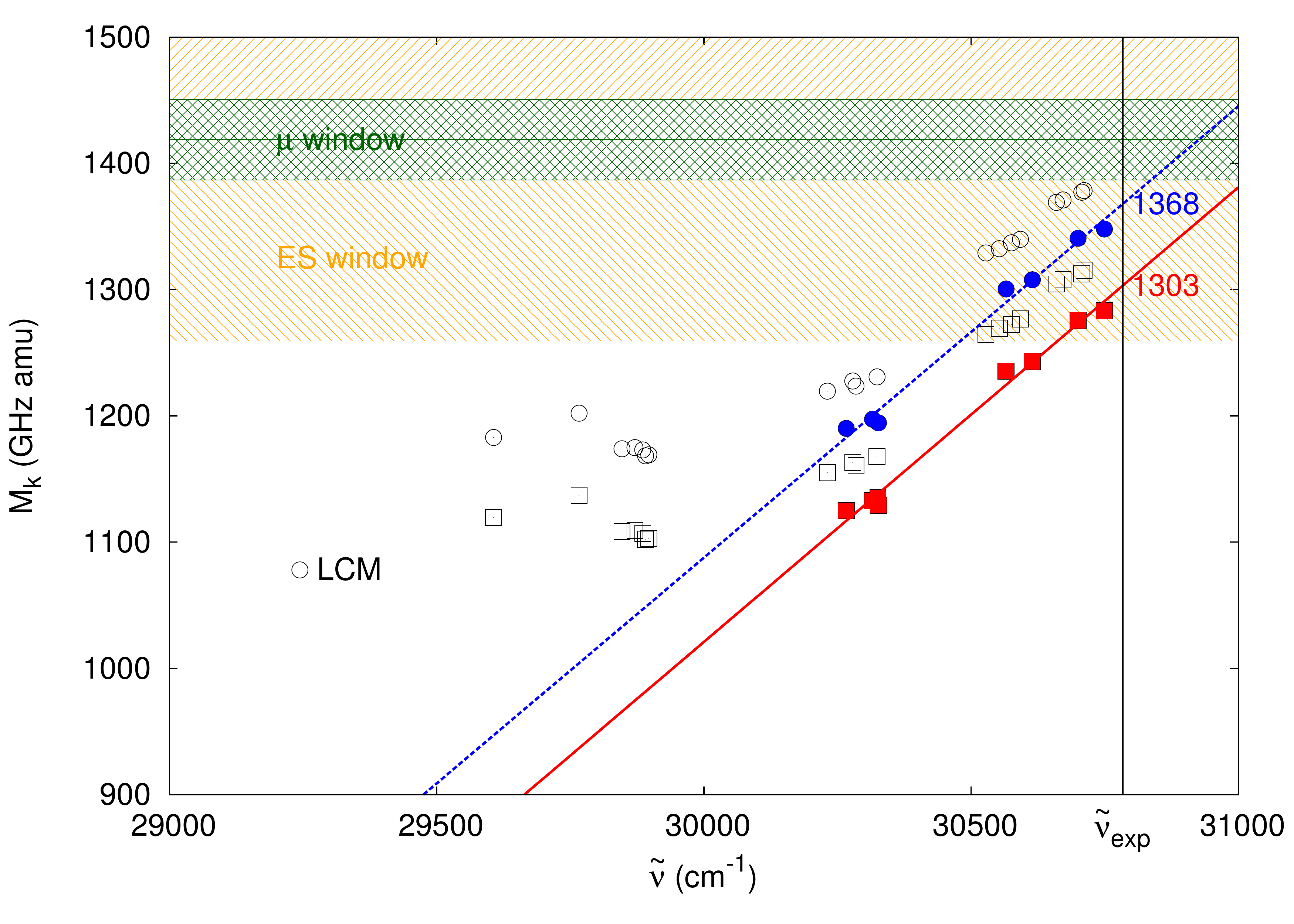}
\caption{(Color online) Mass factor ($M$) versus transition energy ($\tilde{\nu}$) plot for various RCI calculations. The M values allowed by the muonic ($\mu$ window) and electron scattering (ES window) measurements of the nuclear $\delta \langle r^2\rangle $ are shown in green and yellow, respectively. The vertical line indicate the experimental transition energy. Predicted M values obtained with (squares, red) or without (circles, blue) relativistic corrections to the mass shift operators are deduced from extrapolation of results obtained with $n \geq 8$ to the experimental transition energy. \label{fig:conv}}
\end{figure}

It has been observed that  the experimental energy difference between the two atomic levels can provide a good guideline for estimating mass isotope shifts~\cite{Caretal:2010a,CarGod:2011a}. Figure~\ref{fig:conv} shows the calculated mass factor $M_k$ versus the calculated transition energy for a set of RCI calculations, including the effect of inner-core valence correlation as an additive correction. 
The experimental transition energy is indicated as well.  The group of points predicting a transition energy lower than 30~000~cm$^{-1}$ corresponds to the single-reference CV calculations. As $n_{max}$ is increased in these calculations we see that the results converge towards a relatively stable mass factor value around $M_k \sim 1090$ at $\tilde \nu=29~900$~cm$^{-1}$ (convergence starts at $n_{max}=8$).  When including MR-I sets in the calculations, transition energies above 30~000~cm$^{-1}$ are obtained and a linear correlation can be observed between $\tilde \nu$ and $M_k$, as illustrated in Fig.~\ref{fig:conv}. This is consistent with previous observations in robust calculations of isotope shifts on electron affinities~\cite{CarGod:2011a,CarGod:2013a}. It means that, within our model, the error on the computed $M$ values is correlated to the error on the transition energy. Hence we gain in precision by further extrapolating the computed results towards the experimental energy difference, as done in Fig.~\ref{fig:conv}  and reported in Table~\ref{tab:res} under the ``Extrapolated'' entry. From the observed convergence patterns, we deduce an intrinsic error of about 3-4\% on the final value.
As illustrated by the $M~\mbox{(squares)}- \overline{M}~\mbox{(circles)}$ differences, the inclusion of relativistic corrections to the recoil operator brings a significant ($\simeq$~5~\%) correlation-independent reduction of the mass factor. 
\begin{table}
 \center
\caption{Electronic mass and field-shift factors. See text for details on the different calculations. \label{tab:res}}
\begin{tabular}{lcccc}
\toprule
 & \twoline{$\Delta E$\\ $($cm$^{-1})$} & \twoline{$F_k$\\ $($MHz fm$^{-2})$} & \twoline{$\overline{M}_k$\\ $($GHz amu$)$} & \twoline{$M_k$\\$($GHz amu$)$}\\
\midrule
 \hline
 \raisebox{3ex}{}%
DF             & 25~679  & $-$597  &  1166  &  1111\\
\raisebox{4ex}{}%
CC             & 25~744  &  $-$662  &  924   &  862   \\
CC + CV        & 29~892  &   $-$767  &  1169  &  1103  \\
CC + CV + ICV  & 29~892  &   $-$782  &  1156  &  1090 \\
MR-I RCI  & 30~748  &  $-$779  &  1348  &  1283\\
Extrapolated      &         &   & 1368 & 1303 \\
\raisebox{4ex}{}%
Experiment & 30~784 & & &\\
\multicolumn{3}{l}{Electron scattering window (see \cite{Bisetal:2016a})} & \multicolumn{2}{c}{$1258 - 1622$} \\
\multicolumn{3}{l}{Muonic window (see \cite{Bisetal:2016a})} & \multicolumn{2}{c}{$1385 - 1448$} \\
&&&& \\
\bottomrule
\hline
\end{tabular}
\end{table}

Figure~\ref{fig:conv} also includes a range of `semi-empirical' $M_k$-values based on two  data sets, respectively from electron scattering and from muonic atom measurements. 
More details, including the relevant references, can be found in \cite{Bisetal:2016a}. 
Our final extrapolated $M_k$ values fall within the window of semi-empirical $M_k$ values from electron scattering data, but are a few percent below the window from muonic atom data.

Our methodology is unable to provide uncertainties related to types of correlation effects which are neglected. For instance, triple excitations involving electrons among the Argon-like core are not considered. In this context, to take the experimental transition frequency as a reference does not exactly provide the limit of the model.
Assuming a similar correlation between $M_k$ and $\nu_k$ when adding omitted correlation excitations, we have a total additional uncertainty of 5\%. All in all, we see that it is necessary to assume a 5-10\% uncertainty of the final value reported in Table~\ref{tab:res}. With the assumption of this level of uncertainty the calculated $F_k$ and $M_k$ form a completely consistent set when the muonic atom  $\delta\left\langle r^{2}_{\mathrm{c}}\right\rangle^{65,63}$ and observed $\delta\nu^{65,63}$ are considered~\cite{Bisetal:2016a}.

\begin{acknowledgements}
This work has been supported by the BriX IAP Research Programs No. P7/12 (Belgium).\end{acknowledgements}

%

%

\end{document}